\documentclass[10pt,a4paper]{article}
\usepackage{graphicx}
\date{}
\begin{document}


\date{}

\title{Mimicking Dark Matter\\
{\normalsize{Dedicated to Vera Rubin} }}
\author{Ll. Bel\thanks{e-mail:  wtpbedil@lg.ehu.es; www.lluisbel.com}}

\maketitle

\begin{abstract}

I show that a very simple model in the context of Newtonian physics promoted to a first approximation of general relativity can mimic Dark matter and explain most of its intriguing  properties. Namely: i) Dark matter is a halo associated to ordinary matter; ii) Dark matter does not interact with ordinary matter nor with itself; iii) Its influence grows with the size of the aggregate of ordinary matter that is considered, and iv) Dark matter influences the propagation of light.

\end{abstract}

\section{Isolated Point particles}

Let us consider two point particles, one of mass $m$ and the other of mass unity, at a location $x^i$ apart, and assume that Newton's law of attraction is modified so that :

\begin{equation}\label{Newton}
 F^i= -\frac{Gm}{r^3}x^i-Gm\left(\frac{1}{a_1r^2}+\frac{1}{a_2^2r}+\frac{1}{a_3^3}\right)x^i
 \end{equation}
where $x^i$ are Cartesian coordinates, $G$ is Newton's gravitational constant and $a_i$ are three parameters, with dimension length, supposedly to be determined by experiments or observations. It should be understood that at least one of these parameters is not infinite. This formula is a variant of proposals first made in Refs. (\cite{Finzi}), (\cite{Sanders}) and (\cite{Kuhn}).
\,\footnote{In three of my previous arXiv papers, \cite{Bel1}, \cite{Bel2} and \cite{Bel3}, I assumed that $a_2$ and $a_3$ were $\infty$}

Defining the potential $V$, as usual up to an additive constant by:

\begin{equation}\label{Potential}
F_i=-\partial_iV
\end{equation}
we get:

\begin{equation}\label{Laplace}
 \Delta V=4\pi G\rho, \quad \rho=m\delta(r)+\sigma(r), \quad \sigma(r)=\frac{m}{4\pi a_1 r^2} +\frac{m}{2\pi a_2^2 r}+\frac{3m}{2\pi a_3^3 }
\end{equation}
From where it follows that the potential $V$ would be:

\begin{equation}\label{Vequ}
 V=-\frac{G m}{r}+Gm\left(\frac{1}{a_1}\ln(r)+\frac{r}{a_2^2}+\frac{r^2}{a_3^3}\right)
\end{equation}

From the definition of $\sigma$ above it follows that to calculate the force that a particle of mass $m$ exerts on a particle of mass unit it is equivalent to use (\ref{Newton}) or to calculate:

\begin{equation}\label{Sigma}
 F^i=-\frac{G}{r^2}\left(m+4\pi\int_0^u \sigma(u)u^2\, du\right)x^i
\end{equation}
Equivalently we can say that the gravitational mass of the source particle has increased by a fictitious mass, or Dark mass, whose density is $\sigma$.

Thus, dark mass comes to live as halos surrounding every single particle of ordinary mass and we can not expect these halos to interact otherwise with ordinary mass or among themselves if several particles are considered.

Obviously the credibility of this interpretation should be checked against possible contradictions derived from observations.

Several qualitative considerations can be made already. There are in the Universe different aggregates of ordinary matter; e.g., globular clusters (size scale say of the order of $10^2$ ly) , small spheroidal galaxies (size scale say of the order of $10^3$ ly) , large galaxies of different types (size scale say of the order of $10^5$ ly), and aggregates or clusters of them (size scale say of the order of $10^6$ ly). The value of any one of the parameters a's will affect only those structures with a scale larger than the corresponding size scale.If it is large compared to the scale of the parameters $a's$ they will be very much affected by Dark mass, both in the dynamics of its own structure and  in the strength of its gravitational interaction with external objects. If it is small they will  be little affected. This would explain, for instance, that if all three parameters a's are much greater than the size scale of globular clusters, dark matter effects are not important for them, while they are for larger structures. This is nice because it is one of the deep mysteries that everybody has in mente.

\vspace{1cm}
\section{Point particles aggregates}
\vspace{1cm}

A modification of Newton's law of force would modify also the dynamics of Galaxy clusters in dynamical equilibrium through an application of the Virial theorem at an approximation where such clusters could be compared to an aggregate of point particles. Theorem that in this case tells us that:

\begin{equation}\label{Viriel}
 -\frac{G}{2}\sum_{i\neq j}\left(\frac{m_i m_j}{|r_i-r_j|}+m_i m_j\left(\frac{1}{a_1}+\frac{|r_i-r_j|^2}{a_2^2}+\frac{|r_i-r_j|^3}{a_3^3}\right)\right)
+\sum_{i} m_i\left(\frac{dr_i}{dt^2}\right)=0
\end{equation}
contributing thus to explain the anomaly pointed out by Zwicky, (\cite{Zwicky}), who checked that the preceding formula without the terms depending on the a's did  not fit the data.

\vspace{1cm}
\section{Continuous spherical distributions}

Continuous spherical distributions of ordinary matter can be dealt with as usual in Newtonian theory. For example if the density of ordinary matter is a continuous function $\mu(u)$ this will require to replace the function $\sigma(r)$ in (\ref{Laplace}) by:

\begin{equation}\label{sigma}
\sigma=2\pi\int_0^R du\int_0^\pi \frac{\mu(u)u^2\sin\theta}{r^2+u^2-2ru\cos\theta}
\end{equation}
where $R$ is the radius, eventually $\infty$, of the configuration. Considering as an example the cuspidal Hernquist profile:

\begin{equation}\label{Hernquist}
 \mu(u)=\frac{1}{u(1+u)^3}
 \end{equation}
The figures below show the graphics of $\mu(u)$, and $\sigma_a(u)$ corresponding to the parameters $a_1=1$, $a_2=a_3=\infty$ in the first case and $a_2=1$, $a_1=a_3=\infty$ in the second.

\begin{figure}[htbp]
\begin{center}
\begin{minipage}[b]{15cm}
\includegraphics[width=5cm]{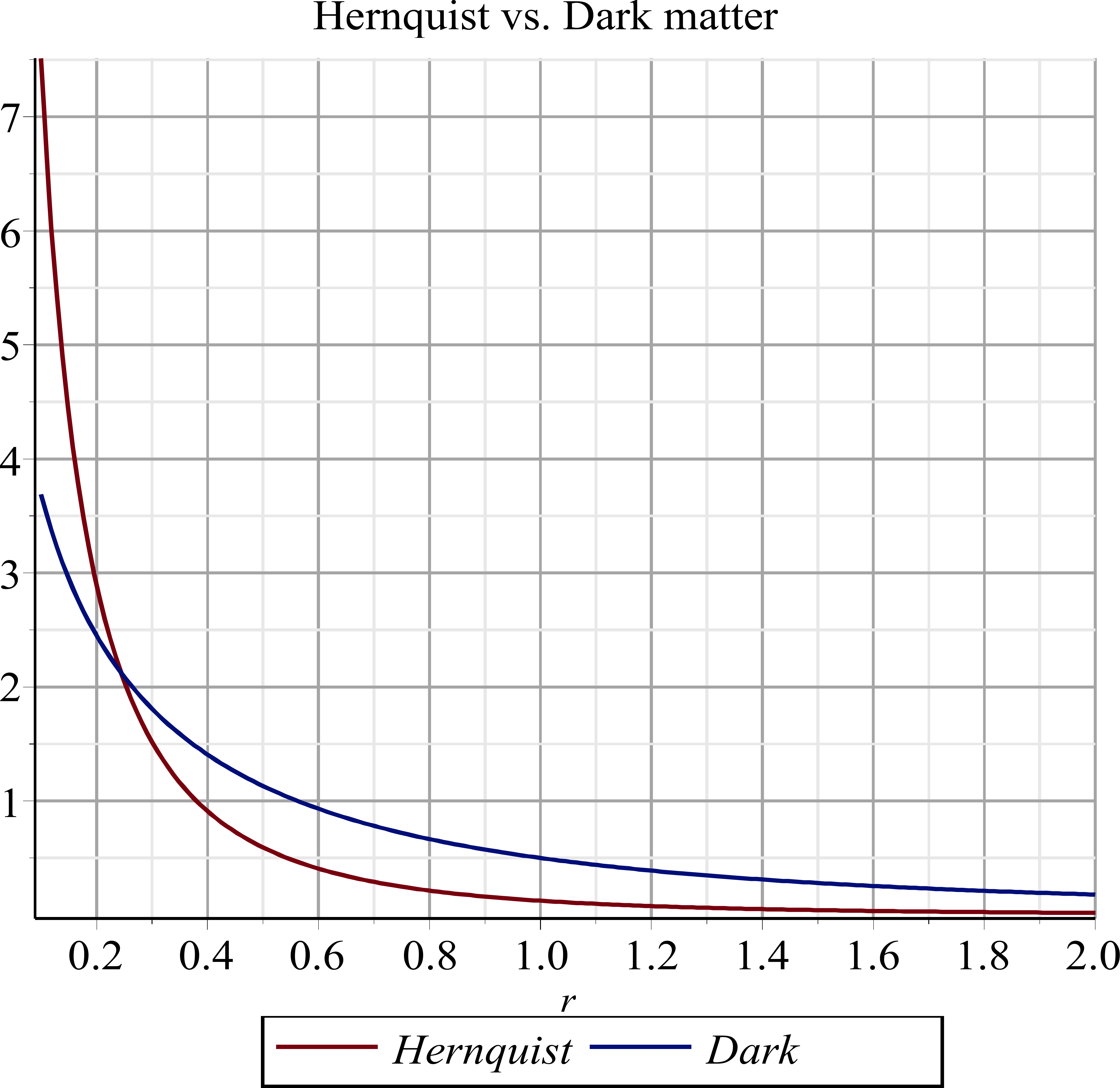}
\hspace{2cm}
\includegraphics[width=5cm]{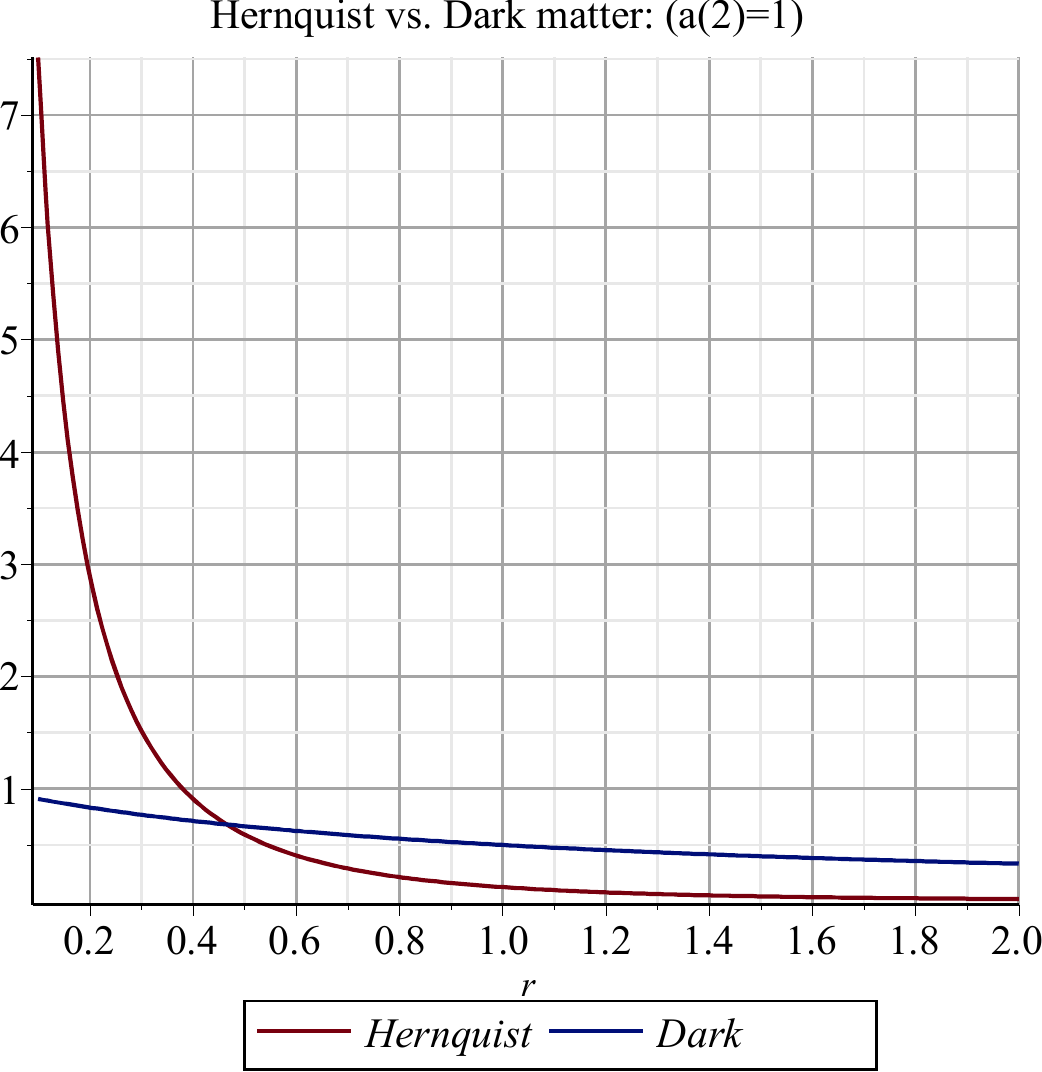}
\end{minipage}
\end{center}
\end{figure}

\section{Rotation curves}
\vspace{1cm}

The familiar Newtonian law for the rotational velocity curves beyond the finite radius of galaxies becomes:

\begin{equation}\label{v-curves}
 v=\sqrt{\frac{Gm}{r}\left(1+\frac{r}{a_1}+\frac{r^2}{a_2^2}+\frac{r^3}{a_3^3}\right)}
\end{equation}

\vspace{1cm}
\section{Light bending }
\vspace{1cm}

There remains to understand how a modification of Newton's law could produce light bending. A problem that I already considered promoting the framework of Newtonian physics to the framework of Einsteín's theory to first order of approximation  in a simplified case: $a_2=a_3=\infty$ (\cite{Bel2}).

I consider here two possible promotions of the Newtonian-like formalism: i) either introducing the line-element:

\begin{equation}
\label{isotropic}
ds^2=-(1+2V(r))dt^2+(1-2V(r))(dr^2+r^2(d\theta^2+\sin^2\theta d\phi^2))
\end{equation}
where $V(r)$ is the potential considered in (\ref{Vequ}). Or ii) the line-element

\begin{equation}\label{Drosde}
ds^2=-(1+2V(r))dt^2+(1-2V(r))dr^2+r^2(d\theta^2+\sin^2\theta d\phi^2)
\end{equation}

Assuming that $a_1=a_2=a_3=\infty$, the line-element (\ref{isotropic}) is the familiar linear approximation of Schwarzschild solution in harmonic or isotropic coordinates, while the line-element (\ref{Drosde}) is the familiar linear approximation of Schwarzschild solution in Drosde's  coordinates. However when the parameters $a's$ have finite values the two line-elements have quite different interpretations that can be made explicit calculating the corresponding Einstein's tensor. In the first case we get:

\begin{equation}\label{EinsteinH}
S_{00}=-Gm\left(\frac{2}{a_1r^2}+\frac{4}{a_2^2r}+\frac{12}{a_3^3}\right)
\end{equation}
the remaining components being zero.

In the second case the non zero components are:

\begin{eqnarray}\label{harmonic}
  S_{00} &=& Gm\left(\frac{2}{r^2a_1}(1+\ln(r)+\frac{4}{ra_2^2}+\frac{6}{a_3^3}\right) \\
  S_{11} &=& Gm\left(-\frac{2}{r^2a_1}(1+\ln(r)-\frac{4}{rb^2}-\frac{6}{a_3^3}\right) \\
  S_{22} &=& Gm\left(-\frac{1}{a_1}-\frac{2r}{a_2^2}-\frac{6r^2}{a_3^3}\right) \\
  S_{33} &=& S_{22}\sin^2\theta
\end{eqnarray}

Notice that (\ref{isotropic}) can be very nicely understood as a solution of Einstein's equation with Dark matter density but no pressure and noteworthy is that assuming $a_1=a_2=\infty$ what we get is that (\ref{EinsteinH}) becomes the first approximation of the Kottler solution (\cite{Kottler}).

On the other hand the interpretation of the line-element (\ref{Drosde}) would require to make sense of the pressure terms.

Now comes a second delicate point. I have used the same letters $r,\theta, \phi$ as if they were polar coordinates of Euclidean space in Newtonian physics. But is this justified on both theories based on the line-elements (\ref{isotropic}) and (\ref{Drosde}). I think not. I think that it is justified in (\ref{isotropic}) and it is not in (\ref{Drosde}) a point that I present more explicitly in the Appendix, and therefore I consider (\ref{isotropic}) to be the proper promotion of the Newtonian generalization that I have been considering.

Now we can deal with light rays as usual in General relativity considering the equations of null auto-parallels. Because of the spherical symmetry of the line-element the space-trajectories of light rays will lie on a plane and it is convenient to assume that this plane is the plane $\phi=0$

The explicit equations are then, $s$ being an arbitrary parameter along the rays:

\begin{eqnarray}
  \frac{d^2 t}{ds^2}-2\frac{dV}{dr}\frac{dr}{ds}\frac{dt}{ds}- b\frac{dt}{ds}=0 \\
  \frac{d^2 r}{ds^2}+\frac{dV}{dr}\left(\frac{dr}{ds}\right)^2 \nonumber \\ -r\left(\frac{dV}{dr}r+1\right)\left(\left(\frac{d\theta}{ds}\right)^2+\left(\frac{d\phi}{ds}\right)^2\right) -\frac{dV}{dr}\left(\frac{dt}{ds}\right)^2 - b\frac{dr}{ds}=0 \\
  \frac{d^2 \theta}{ds^2}+\frac{2}{r}\left(\frac{dV}{dr}r+1\right)\frac{dr}{ds}\frac{d\theta}{ds} = 0 \\
  \frac{d^2 \phi}{ds^2}+\frac{2}{r}\left(\frac{dV}{dr}r+1\right)\frac{dr}{ds}\frac{d\phi}{ds} - b\frac{d\theta}{ds}
\end{eqnarray}
$b$ being a function of $s$, that is by definition an affine parameter if $b=0$.

Choosing as parameter $s=t$ reduces the preceding system to the following two equations:

\begin{eqnarray}
\label{Bending1}
\frac{d^2r}{dt^2}+\frac{dV}{dr}\left(3\left(\frac{dr}{dt}\right)^2-r^2\left(\frac{d\theta}{dt}\right)^2-1\right)-r\left(\frac{d\theta}{dt}^2\right) &=& 0 \\
\label{Bending2}
\frac{d^2\theta}{dt^2}+\frac{d\theta}{dt}\frac{dr}{dt}\left(4\frac{dV}{dr}+\frac{2}{r}\right) &=& 0
\end{eqnarray}
That it is equivalent to the system of equations of the geodesics of the three dimensional elliptic metric:

\begin{equation}\label{dt2}
 dt^2=(1-4V)(dr^2+r^2(d\theta^2+\sin^2\theta\,d\theta^2)
\end{equation}
These equations have to be integrated using initial conditions:

\begin{equation}\label{IC}
r_0,\ \left(\frac{dr}{dt}\right)_0,\ \theta_0,\ \left(\frac{d\theta}{dt}\right)_0
\end{equation}
satisfying the following condition:

\begin{equation}
    -(1+2V(r_0))+(1-2V(r_0))\left(\left(\frac{dr}{dt}\right)_0^2+r_0^2\left(\frac{d\theta}{dt}\right)_0^2\right)=0.
\end{equation}

Let us now calculate the general relativistic bending of light by the Sun assuming that $a_1=a_2=a_3=\infty$. Using units such that $G=c=r_0=1$,  $r_0$ being the radius of the Sun it turns out that the mass of the Sun is $m=0.000002136477017$ meters. Integrating the system of differential equations (\ref{Bending1}) and (\ref{Bending2}) with initial conditions:

\begin{equation}\label{Initial conditions}
r_0=1 ,\ \theta_0=0,\ \left(\frac{dr}{dt}\right)_0=0,\ r_0\left(\frac{d\theta}{dt}\right)_0=1+\frac{2m}{r_0}
\end{equation}
from $t=-t_\infty$ to $t=+t_\infty$, $t_\infty$ being a sufficiently large value of $t$.
we get:

\begin{equation}\label{Delta}
 \Delta\theta=0.00000875\,rad
\end{equation}
that correspond to 1.8 arcsec,
which is the well known result\,\footnote{the reference \cite{Brown} gives a list of historical results}. It is easy to check repeating the integration process with three set of parameters: i) $a_1=100, a_2=\infty, a_3=\infty$, ii) $a_1=\infty, a_2=100, a_3=\infty)$ and iii) $a_1=\infty, a_2=\infty, a_3=1000$ the corresponding bending results are equal to the observed observed one. Therefore these parameters or greater ones can not be excluded to mimick dark matter beyond the solar system to much larger scales.

\section*{Appendix}

Let us consider a general relativistic metric:

\begin{equation}\label{ds2}
ds^2=-A^2(t,x^k)(-dt^2+f_i(t,x^k)dx^i)^2+A^{-2}\bar g_{ij}(t,x^k)dx^idx^j, \quad i,j,k=1,2,3
\end{equation}
Call the time-like world-lines congruence $t$ variable the reference system, and the 3-dimensional metric $\bar g_{ij}$ the metric of space. Consider now the Euclidean metric (or any other 3-dimensional constant curvature metric):

\begin{equation}\label{Euclidean}
 d\tilde s^2=\tilde g_{ij}(t,x^k)dx^idx^j
\end{equation}
where $x^k$ are whatever coordinates suits you. Then its meaning in (\ref{Euclidean}) and in (\ref{ds2}) will be the same if and only if:

\begin{equation}\label{Meaning}
(\bar\Gamma^i_{jk}-\tilde\Gamma^i_{jk})\bar g^{jk}=0
\end{equation}
where the $\Gamma's$ are the Christoffel connection symbols of $\bar g_{ij}$ and $\tilde g_{ij}$

\vspace{1cm}
{\bf Acknowledgments}

Juan Mari Aguirregabiria and Luis Acedo helped me to improve this manuscript.


\begin{thebibliography}{9}

\bibitem{Zwicky} F.\ Zwicky, Helv.Phys.Acta, 6, 1933
\bibitem{Finzi} A.\ Finzi, MNRAS 127, 1963
\bibitem{Sanders}, R.\ H.\ Sanders, Astron. Astrophys, 136, 1984
\bibitem{Kuhn} J.\ R.\ Kuhn \&
 L.\ Kruglyak, Ap. J., 1987
\bibitem{Bel1} Ll. Bel,  arXiv:1308.0249v2 [physics.gen-ph]
\bibitem{Bel2} Ll. Bel,  arXiv:1311.6891v1[gr-qc]
\bibitem{Bel3} Ll. Bel,  arXiv:1404.0225v1[gr-qc]
\bibitem{Kottler} F. Kottler, Annalen Physic, 410 (1918)
\bibitem{Brown} K. Brown {\it Reflections on Relativity}, page 422, 2016

\end{thebibliography}
\end{document}